\documentclass[twocolumn,showpacs,aps]{revtex4}
\usepackage{graphicx}
\usepackage{bm}
\usepackage{amsmath}
\usepackage{amssymb}
\usepackage{color}

\def\l{\langle}
\def\r{\rangle}

\begin{document}
\title{
Interplay of dilution and magnetic field in the nearest-neighbor 
spin-ice model \\
on the pyrochlore lattice
} 
\author{Alexey Peretyatko$^{1}$}
\email{peretiatko.aa@dvfu.ru}
\author{Konstantin Nefedev$^{1,2}$}
\email{nefedev.kv@dvfu.ru}
\author{Yutaka Okabe$^{3}$}
\email{okabe@phys.se.tmu.ac.jp}
\affiliation{
$^1$School of Natural Sciences, Far Eastern Federal University, Vladivostok, 
Russian Federation \\
$^2$Institute of Applied Mathematics, Far Eastern Branch, 
Russian Academy of Science, Vladivostok, Russian Federation \\
$^3$Department of Physics, Tokyo Metropolitan University, Hachioji, 
Tokyo 192-0397, Japan \\
}

\date{\today}

\begin{abstract}
We study the magnetic field effects on the diluted spin-ice materials 
using the replica-exchange Monte Carlo simulation. 
We observe {\it five} plateaus in the magnetization curve of the diluted 
nearest-neighbor spin-ice model on the pyrochlore lattice 
when a magnetic field is applied in the [111] direction. 
This is in contrast to the case of the pure model with 
two plateaus.  The origin of {\it five} plateaus is investigated 
from the spin configuration of two corner-sharing tetrahedra 
in the case of the diluted model. 
\end{abstract}

\pacs{
75.40.Mg, 75.50.Lk, 64.60.De
}

\maketitle


Spin-ice materials have captured a lot of attention 
\cite{Harris,Ramirez,Bramwell}, and their exotic physics 
is a current topic of geometrically frustrated magnets. 
Prototype materials are pyrochlores Dy${_2}$Ti${_2}$O${_7}$ and 
Ho${_2}$Ti${_2}$O${_7}$. 
Magnetic field effects, especially the existence of plateaus, 
have been studied theoretically \cite{Harris98,Moessner,Isakov} 
and experimentally 
\cite{Matsuhira02,Hiroi,Sakakibara,Higashinaka,Fukazawa}. 

In spin-ice materials, magnetic ions (Dy$^{3+}$ 
or Ho$^{3+}$) occupy the pyrochlore lattice of corner-sharing 
tetrahedra, and the local crystal field environment causes 
magnetic moments to point along the lines connecting 
the centers of the two tetrahedra at low temperatures. 
In the low-temperature spin-ice state, 
magnetic moments are highly constrained locally and 
obey the so-called ``ice rules": two spins point in and 
two spins point out of each tetrahedron of 
the pyrochlore lattice. This 2-in 2-out spin configuration 
is equivalent to the situation for hydrogen atoms in water ice 
\cite{Pauling}.

The dilution effect on frustration was studied by Ke {\it et al.} 
\cite{Ke} for spin-ice materials.  Magnetic ions 
Dy or Ho are replaced by nonmagnetic Y ions. 
Nonmonotonic zero-point entropy as a function of the dilute concentration 
was observed experimentally, and further studies on the dilution 
effects have been reported \cite{Lin,Scharffe,Shevchenko}.

In this communication, we study the diluted nearest-neighbor (NN)
antiferromagnetic (AFM) Ising model on the pyrochlore lattice 
under a magnetic field in the [111] direction. 
We treat the NN interaction as a theoretical model; 
a more complicated model, such as the dipolar model, may be required to make
connections to actual materials.
For the simulation method, we use the replica-exchange Monte Carlo 
method \cite{Hukushima} to avoid the trap at local-minimum 
configurations. 



We are concerned with the Hamiltonian, which is given 
in Eq.~(2.2) of Ref.~\cite{Isakov}; 
\begin{equation}
 H = J \sum_{\l i,j \r} \sigma_i \sigma_j
   - \sum_i \bm{h} \cdot \bm{d}_{\kappa(i)} \ \sigma_i, 
\end{equation}
where $J (>0)$ is the effective antiferromagnetic coupling, 
$\sigma_i$ are the Ising pseudospins ($\sigma_i = \pm 1$), and 
$\l i,j \r$ stands for the NN pairs. 
The unit vectors $\bm{d}_{\kappa(i)}$ are the local easy axes 
of the pyrochlore lattice, and explicitly described as 
$
\bm{d}_{\kappa(i)} = \{ \bm{d}_0, \bm{d}_1, \bm{d}_2, \bm{d}_3 \}, 
$
where
$\bm{d}_0 = (1,1,1)/\sqrt{3}$, 
$\bm{d}_1 = (1,-1,-1)/\sqrt{3}$, 
$\bm{d}_2 = (-1,1,-1)/\sqrt{3}$, 
and 
$\bm{d}_3 = (-1,-1,1)/\sqrt{3}$.
We consider the case when the magnetic field $\bm{h}$ is along 
the [111] direction; that is,
$
   \bm{h} = h \bm{d}_0.
$
Then, $\bm{h} \cdot \bm{d}_{\kappa(i)}$ becomes $h$ 
for apical spins where $\bm{d}_{\kappa(i)} = \bm{d}_0$, 
but $-(1/3)h$ for other spins. 
The magnetization $M$ along the [111] direction is 
calculated through the relation
\begin{equation}
   M = \sum_i \bm{d}_0 \cdot \bm{d}_{\kappa(i)} \ \sigma_i.
\end{equation}
%

\begin{figure}[t]
\begin{center}
\includegraphics[width=6.0cm]{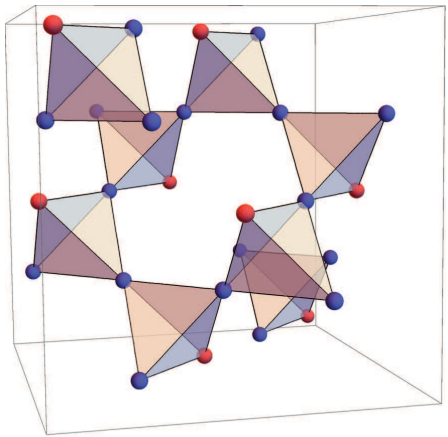}
\caption{
(Color online.)
The illustration of the pyrochlore lattice, which is a three-dimensional 
network of corner-sharing tetrahedra.  The apical spins, 
$\bm{d}_{\kappa(i)} = \bm{d}_0$, are shown in red, 
whereas other spins are in blue.
}
\label{fig:pyro}
\end{center}
\end{figure}

For convenience, the illustration of the pyrochlore lattice, 
which is a three-dimensional network of corner-sharing tetrahedra, 
is given in Fig.~\ref{fig:pyro}, where the apical spins, 
$\bm{d}_{\kappa(i)} = \bm{d}_0$, are shown in red, 
whereas other spins are in blue.

In the case of the site dilution of spins, the Hamiltonian becomes
\begin{equation}
 H = J \sum_{\l i,j \r} c_i c_j \sigma_i \sigma_j
   - \sum_i \bm{h} \cdot \bm{d}_{\kappa(i)} \ c_i \sigma_i. 
\end{equation}
Here $c_i$ are the quenched variables ($c_i = 1 \ {\rm or} \ 0$), 
and the concentration of vacancies is denoted by $x$. 


The Monte Carlo simulation has become a standard method 
to study many-body problems in physics. However, we sometimes 
suffer from the problem of slow dynamics.  
One attempt to solve the problem of slow dynamics 
is the extended ensemble method. 
The replica-exchange Monte Carlo method or 
parallel tempering \cite{Hukushima,Marinari} 
is an example.
The replica-exchange method has been successfully applied 
to various systems, for example, in the problem of 
spin glasses \cite{Lee}. The idea of replica-exchange can be 
combined with the molecular dynamics simulation. 
Using this combined method, the protein-folding problem 
was studied \cite{Sugita}. 

Let us consider two replicas of a system. 
If the inverse temperature of the replica 1 is 
$\beta_1$ and that of the replica 2 is $\beta_2$, 
the Boltzmann weight of the composite system is 
$\exp[-(\beta_1 E_1 + \beta_2 E_2)]$, where $E_1$ and $E_2$ 
are the energy of each system.  We try to exchange two replica 
systems, or exchange two inverse temperatures. 
If we make this exchange trial using the transition 
probability based on the relative Boltzmann weight,
\begin{eqnarray*}
  &~&\exp[-(\beta_1 E_2+\beta_2 E_1)+(\beta_1 E_1+\beta_2 E_2)] \\
   &~&\hspace{1cm} =\exp[(\beta_1-\beta_2)(E_1-E_2)],
\end{eqnarray*}
the Boltzmann distribution of the composite system is guaranteed.
In the actual calculation, we may use many replicas. 
We note that replicas are not necessarily for several 
temperatures; we can also use replicas for several magnetic fields. 
In the present simulation, we treat 546 systems of 
91 magnetic fields and 6 temperatures simultaneously. 
To escape from the local-minimum trap problem, 
the loop algorithm is sometimes employed \cite{Isakov}. 
Here, we use the replica-exchange method for both temperature 
and magnetic field based on the single spin flip. 
The convergence is good enough even for low temperatures 
such as $T/J=0.05$.

For the simulation of the AFM Ising model on the pyrochlore lattice, 
we used a 16-site cubic unit cell of the pyrochlore lattice, 
and systems with $L \times L \times L$ 
unit cells with periodic boundary conditions were treated. 
We performed simulations for the system sizes of $L$ =6, 8, and 10; 
the numbers of sites were $N$ = 3456, 8192, and 16000, 
respectively.
As for the dilute concentration $x$, we treated 
$x$= 0.0 (pure), 0.2, 0.4, 0.6, and 0.8. 
We discarded the first 5,000 Monte Carlo Steps (MCSs) 
to avoid the effects of initial configurations, 
and the next 50,000 MCSs were used for the measurement.
We took an average over 20 samples for each size 
and each $x$ to estimate the statistical errors. 


\begin{figure}
\begin{center}
\includegraphics[width=8.0cm]{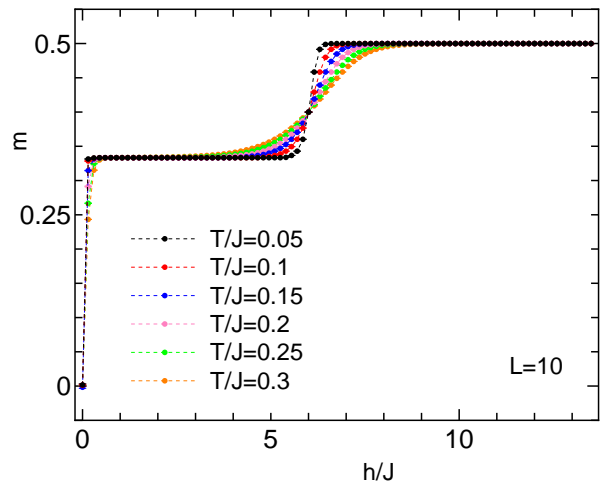}
\caption{
(Color online) 
The magnetization curve for the AFM Ising model 
on the pyrochlore lattice. 
The magnetic field is applied in the [111] direction. 
The system size is $L=10 \ (N=16000)$, 
and the temperature is $T/J$ = 0.05, 0.1, 0.15, 0.2, 0.25, and 0.3. 
}
\label{fig:pure_mag}
\end{center}
\end{figure}

We start with showing the results of the pure (non-diluted) model ($x=0.0$). 
In Fig.~\ref{fig:pure_mag} the magnetization $m=M/N$ is 
plotted as a function of the applied field $h$. 
The system size is $L=10 \ (N=16000)$. 
We plot the data for the temperatures $T/J$ = 0.05, 0.1, 0.15, 
0.2, 0.25, and 0.3. 
The average was taken over 20 samples with different random-number 
sequences. 
The statistical errors are within the size of the marks.
The size $L=10$ is large enough; the size dependence is small. 
We confirmed the two plateaus in the magnetization curve; 
in the first of which the ground-state entropy is reduced 
but still remains extensive \cite{Moessner,Isakov}.
The height of the plateau jumps from $m=1/3$ to $m=1/2$ 
at $h/J=6$, and the jump becomes smoother 
when raising the temperature.
The pyrochlore lattice can be regarded as alternating kagome 
and triangular layers, and the [111] magnetic field effectively 
decouples these layers. We can see in Fig.~1 that blue spins form 
the kagome lattice.  The states with $m=1/3$ still have macroscopic 
degeneracy, and sometimes called as "kagome-ice" states 
\cite{Hiroi,Sakakibara}. 
The magnetization of $m=1/2$ is the maximum magnetization for this problem. 

\begin{figure}
\begin{center}
\includegraphics[width=8.0cm]{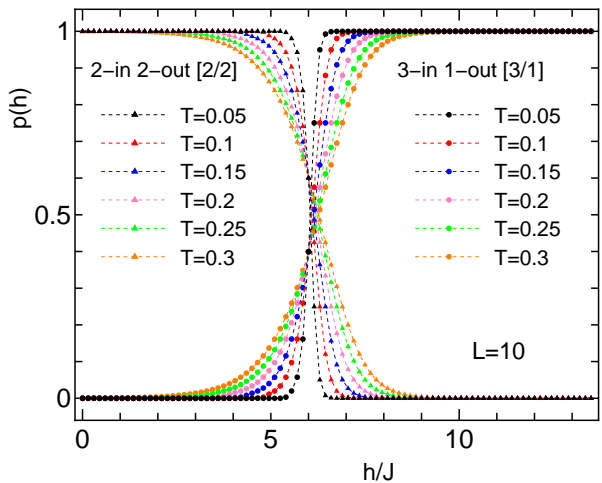}
\caption{
(Color online) 
The proportion of the types of spin configurations in the tetrahedron 
for the AFM Ising model on the pyrochlore lattice 
under the magnetic field. 
The magnetic field is applied in the [111] direction. 
The system size is $L=10 \ (N=16000)$, 
and the temperature is $T/J$ = 0.05, 0.1, 0.15, 0.2, 0.25, and 0.3. 
}
\label{fig:pure_config}
\end{center}
\end{figure}

The proportion of the types of spin configurations in the tetrahedron 
for the AFM Ising model on the pyrochlore lattice 
under the magnetic field in the [111] direction 
is plotted in Fig.~\ref{fig:pure_config}. 
There are 8000 tetrahedra for $L=10$, and the proportion of 
the types of spin configurations was measured for 50,000 MCSs. 
An average over 20 independent 
samples to estimate the error bars.
The type of spin configurations is classified by 
the numbers of up ($+1$) spins and those of down ($-1$) spins 
in the tetrahedron.  The up spin points "out" in one tetrahedron, 
but it points "in" in the adjacent tetrahedron. 
Here, we use the notation "out" for up spin and 
the notation "in" for down spin 
to express a spin configuration.
We clearly see the change from the 2-in 2-out 
configuration to the 3-in 1-out configuration 
at $h/J=6$. The change becomes smoother 
when the temperature is raised.


\begin{figure}
\begin{center}
\includegraphics[width=8.0cm]{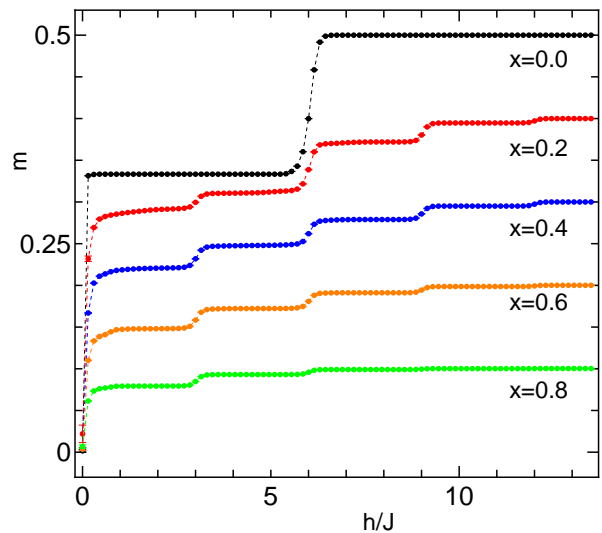}
\caption{
(Color online) 
The magnetization curve for the diluted AFM Ising model 
on the pyrochlore lattice. The system size is $L=10 \ (N=16000)$, 
and the temperature is $T/J=0.05$. 
The dilute concentration ($x$) is 0.0, 0.2, 0.4, 0.6, and 0.8.
}
\label{fig:dilution_mag}
\end{center}
\end{figure}

Next, we consider the results of diluted systems. 
We plot the magnetization curve for the diluted AFM Ising model 
on the pyrochlore lattice in Fig.~\ref{fig:dilution_mag}. The system size is $L=10 \ (N=16000)$, 
and the temperature is $T/J=0.05$. 
The dilute concentration ($x$) is 0.0, 0.2, 0.4, 0.6, and 0.8.
The average was taken over 20 random samples. 
The error bars are shown in the figure, but they are within 
the size of marks. 
The random average over 20 samples yields small statistical errors 
for large enough system size of $L=10 \ (N=16000)$.

We observe five plateaus in the magnetization curve of diluted 
systems; 
for $h/J<3$, $3<h/J<6$, $6<h/J<9$, $9<h/J<12$, and $h/J>12$. 
This is in contrast to the pure case where only 
two plateaus are separated at $h/J=6$.
We plot the results of $T/J=0.05$; if the temperature is 
raised, the magnetization step becomes smoother.
The saturated value of the magnetization $m$ 
is $(1/2)*(1-x)$.

\begin{figure}
\begin{center}
\includegraphics[width=8.2cm]{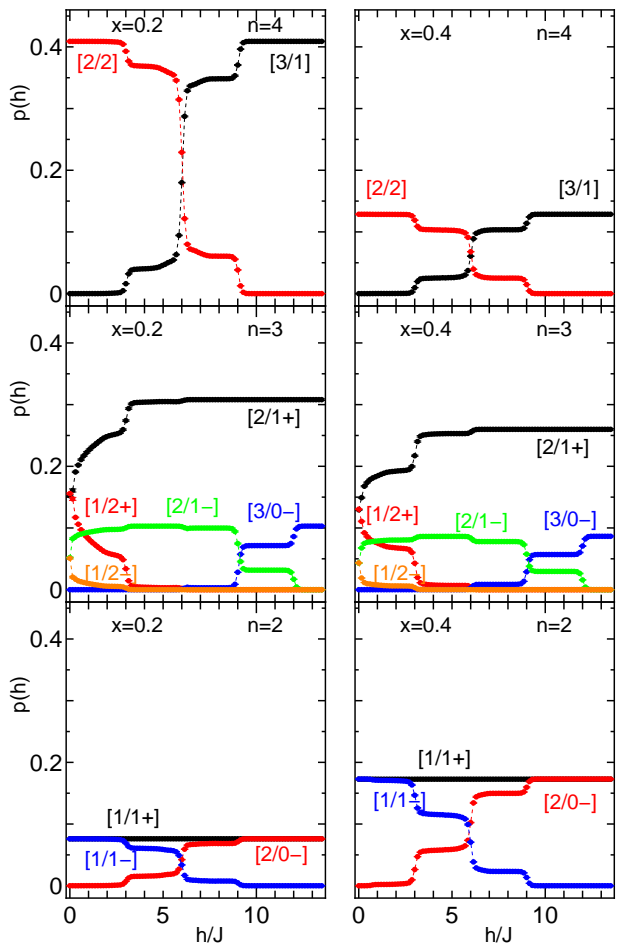}
\caption{
(Color online) 
The proportion of the types of spin configurations in the tetrahedron 
for the diluted AFM Ising model on the pyrochlore lattice 
under the magnetic field in the [111] direction. 
The system size is $L=10 \ (N=16000)$, 
and the temperature is $T/J$ = 0.05. 
The dilute concentration ($x$) is 0.2 (left) and 0.4 (right).
The number of spins $n$ in the tetrahedron is 4, 3, and 2 
for the top, middle, and bottom panel, respectively. 
}
\label{fig:dilution_config}
\end{center}
\end{figure}

In the case of the pure model, the spin configuration of the tetrahedron 
changes from the 2-in 2-out configuration to the 3-in 1-out configuration 
when the magnetic field is applied in the [111] direction. 
On the contrary, the spin configuration becomes more complex 
for diluted systems.  The situation becomes different 
if an apical spin is deleted or another spin is deleted. 
The magnetic field dependence of the spin configuration 
is plotted in Fig.~\ref{fig:dilution_config}.
The system size is $L=10 \ (N=16000)$, 
and the temperature is $T/J$ = 0.05. 
The dilute concentration ($x$) is 0.2 and 0.4.
There are 8000 tetrahedra for $L=10$, 
and the number of spins in the tetrahedron, $n$, becomes 
4, 3, 2, 1, or 0 for diluted systems.

The top panel of Fig.~\ref{fig:dilution_config} is 
the plot for the spin number $n=4$. 
The number of spins, $n$, is 4 for about 41\% of tetrahedra 
in the case of $x=0.2$, and about 13\% in the case of $x=0.4$.
The change from the 2-in 2-out configuration, [2/2], to 
the 3-in 1-out configuration, [3/1], is observed 
at $h/J=6$ as in the pure case.  
However, the proportions of [2/2] and [3/1] also change 
at $h/J=3$ and at $h/J=9$. 

The middle panel of Fig.~\ref{fig:dilution_config} is 
the plot for the spin number $n=3$. 
There are two possibilities that the deleted spin 
is an apical spin and that is another spin. 
When the apical spin is not deleted, the partial change from 
the 1-in 2-out configuration, [1/2+], to 
the 2-in 1-out configuration, [2/1+], is observed. 
When the apical spin is deleted, 
the change from the 2-in 1-out configuration, $[2/1-]$, to
the 3-in 0-out configuration, $[3/0-]$, is observed with two steps 
at $h/J=9$ and at $h/J=12$. 
For the low-$h$ region, the 1-in 2-out configuration, $[1/2-]$, 
remains. We use the notations "$+$" and "$-$" to specify 
the configuration whether the apical spin is deleted ("$-$") 
or not deleted ("$+$"). Thus, [2/1+], for example, stands for 
the 2-in 1-out configuration where there are three spins including 
the apical spin in the tetrahedron.

The bottom panel of Fig.~\ref{fig:dilution_config} is 
the plot for the spin number $n=2$. 
Two spins are deleted from the tetrahedron. 
If the apical spin is not deleted, 
the 1-in 1-out configuration, [1/1+], is stable.
When one apical spin is deleted, 
the change from the 1-in 1-out configuration, $[1/1-]$, to
the 2-in 0-out configuration, $[2/0-]$, at $h/J=6$ is observed. 
The variation in the proportions 
is observed at $h/J=3$ and $h/J=9$.

We showed the data for $x=0.2$ and $0.4$, and we see 
the dilute concentration ($x$) dependence. 
The situation is essentially the same, 
although the proportion of smaller $n$ in the tetrahedron 
increases when $x$ becomes larger.  For strong dilution, 
such as $x=0.8$, there appear many free spins, 
which do not create a magnetization step; 
the magnetization jump becomes smaller for larger $x$. 
Even the proportion of tetrahedra with no spins ($n=0$) 
increases, which do not have magnetization.


\begin{table}
\caption{
The local energy of the spin configuration in the tetrahedron 
for the spin numbers $n$= 4, 3, and 2. 
When the magnetic field is applied in the [111] direction, 
the apical spin is fixed as "out". 
}
\label{configuration}
\begin{center}
\begin{tabular}{llllll}
\hline
\hline
config. & $n$ spins & apical ("out") & "in" & "out" & energy \\
\hline
$[3/1]$  & \ 4 & \ 1 & \ 3 & \ 0 & \ $-6(h/6)$ \\
$[2/2]$  & \   & \ 1 & \ 2 & \ 1 & \ $-2J-4(h/6)$ \\
$[2/1+]$ & \ 3 & \ 1 & \ 2 & \ 0 & \ $-J-5(h/6)$ \\
$[1/2+]$ & \   & \ 1 & \ 1 & \ 1 & \ $-J-3(h/6)$ \\
$[3/0-]$ & \   & \ 0 & \ 3 & \ 0 & \ $3J-3(h/6)$ \\
$[2/1-]$ & \   & \ 0 & \ 2 & \ 1 & \ $-J-(h/6)$ \\
$[1/2-]$ & \   & \ 0 & \ 1 & \ 2 & \ $-J+(h/6)$ \\
$[1/1+]$ & \ 2 & \ 1 & \ 1 & \ 0 & \ $-J-4(h/6)$ \\
$[2/0-]$ & \   & \ 0 & \ 2 & \ 0 & \ $J-(h/6)$ \\
$[1/1-]$ & \   & \ 0 & \ 1 & \ 1 & \ $-J$ \\
\hline
\end{tabular}
\end{center}
\end{table}

Now, we elucidate the origin of the five plateaus 
in the magnetization curve.  
We investigate the local energy of the spin configuration of tetrahedron 
for spin numbers $n$= 4, 3, and 2, which is 
tabulated in Table \ref{configuration}. 
The apical spin is fixed as "out" when the magnetic field is 
applied in the [111] direction. 
The sum of "in" and "out" spins of other spins is $n-1$ when the 
apical spin is not deleted, whereas that is $n$ 
when the apical spin is deleted. 
The local energy for each configuration is given in the last column. 
We note that the Zeeman energy term is shared by two tetrahedra. 

\begin{figure}[t]
\begin{center}
\includegraphics[width=8.4cm]{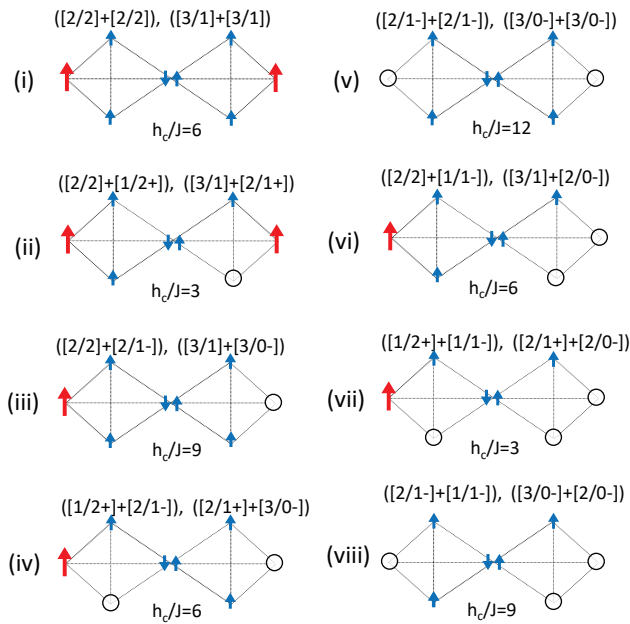}
\caption{
(Color online) 
The schematic illustration of spin flip for corner-sharing 
tetrahedra.
The apical spins are denoted by the long red arrow, 
whereas other spins are denoted by the short blue arrow. 
The deleted spins are denoted by the empty circle.
The cases that two tetrahedra are $n=4, 3, 2$ are 
shown in (i) $\sim$ (viii). The crossover values $h_c$ 
are given there.
}
\label{fig:dilution_flip}
\end{center}
\end{figure}

We consider the energy of two corner-sharing tetrahedra. 
In Fig.~\ref{fig:dilution_flip}, the flip process is 
schematically illustrated. 
The apical spins are denoted by the long red arrow, 
whereas other spins are denoted by the short blue arrow. 
We note that $\bm{d}_0 \cdot \bm{d}_{\kappa(i)}$ = 1 for 
the apical spin, whereas that is $-1/3$ for other spins.
The deleted spins are denoted by the empty circle.
The case when the two tetrahedra are $n=4$ is shown in (i). 
When the corner-sharing spin is turned from "out" to "in", 
the configuration is changed from ([2/2]+[2/2]) to ([3/1]+[3/1]). 
From Table \ref{configuration} the crossover magnetic field 
is calculated by
\begin{eqnarray*}
 &~& 2(-2J-4(h/6)) = 2(-6(h/6)). 
\end{eqnarray*}
Then, we get $h_c/J=6$.
The cases when the two tetrahedra are $n=4$ and $n=3$ are shown in (ii) 
and (iii). 
When the corner-sharing spin is turned from "out" to "in", 
the configuration is changed from ([2/2]+[1/2+]) to 
([3/1]+[2/1+]) in (ii). 
From Table \ref{configuration} the crossover magnetic field 
is calculated by
\begin{eqnarray*}
 &~&(-2J-4(h/6))+(-J-3(h/6)) \\
 &~&\hspace{1cm} =(-6(h/6))+(-J-5(h/6)). 
\end{eqnarray*}
Then, we get $h_c/J=3$.  A similar consideration 
yields $h_c/J=9$ for the change from ([2/2]+$[2/1-]$) to 
([3/1]+$[3/0-]$) in (iii) from the relation 
\begin{eqnarray*}
 &~&(-2J-4(h/6))+(-J-(h/6)) \\
 &~&\hspace{1cm} =(-6(h/6))+(3J-3(h/6)). 
\end{eqnarray*}
The cases when both are $n=3$ are shown in (iv) and (v). 
The case that two tetrahedra are $n=4$ and $n=2$ is shown in (vi). 
The cases when they are $n=3$ and $n=2$ are shown in (vii) 
and (viii).  The crossover values $h_c$ can be obtained 
in the same way as before, and they are given 
in Fig.~\ref{fig:dilution_flip}.

This investigation clearly accounts for the change in the 
configurations given in Fig.~\ref{fig:dilution_config}, 
which leads to the elucidation of the {\it five} 
magnetization plateaus shown in Fig.~\ref{fig:dilution_mag}.


To summarize, we studied the magnetic field effects 
on diluted spin-ice materials 
using the replica-exchange Monte Carlo simulation. 
We observed {\it five} plateaus in the magnetization curve 
of the diluted NN spin-ice model 
on the pyrochlore lattice when a magnetic field 
was applied in the [111] direction. 
This is in contrast to the case of the pure model with 
two plateaus.  
The origin of the {\it five} plateaus was investigated 
from the spin configuration of two corner-sharing tetrahedra 
in the case of the diluted model. 

As for the actual material, the magnetization step was 
observed around 0.9 Tesla for Dy${_2}$Ti${_2}$O${_7} $\cite{Sakakibara,Kadowaki}.
In the diluted system, more steps could be observed 
at the half magnetic field ($h/J=3$) of the pure system ($h/J=6$) 
and at the one-and-a-half magnetic field ($h/J=9$).  
These two new steps are rather easy to be observed
because these steps appear when a single spin 
is deleted from two corner-sharing tetrahedra 
(See (ii) and (iii) in Fig.~\ref{fig:dilution_flip}). 
On the other hand, the step at $h/J=12$ is smaller 
because this happens only when the two spins in the adjacent 
tetrahedra are deleted (See (v) 
in Fig.~\ref{fig:dilution_flip}), 
although the change of spin configuration is observed at $h/J=12$ 
as in the middle panel of Fig.~\ref{fig:dilution_config} 
even for large $x$. 
We treated the diluted NN spin-ice model. 
The long-range dipolar interaction may have some effects, 
although the important point is the interplay of dilution 
and magnetic field for frustrated systems. 
The competition between the pair interaction term and the Zeeman 
term becomes complex when spins are deleted. 
Experimental researches are awaited. 

Antiferromagnetic spin systems on the pyroclore lattice 
provide a rich variety of physics of 
frustrated systems.  In the present communication 
we discussed the interplay of dilution and magnetic field.
The relevance to the magnetic monopoles picture \cite{Castelnovo,Kadowaki} 
will be an interesting problem.
The effects of magnetic fields in other directions 
considering the Kasteleyn transition 
\cite{Kasteleyn} is left to a future work.

\smallskip

We thank Vitalii Kapitan, Yuriy Shevchenko, and 
Konstantin Soldatov for valuable discussions. 
The computer cluster of Far Eastern Federal University 
was used for computation.
This work was supported by a Grant-in-Aid for Scientific Research 
from the Japan Society for the Promotion of Science,  
Grant Numbers JP25400406, JP16K05480.

\end{document}